# Near-field effects on cathodoluminescence outcoupling in perovskite thin films

Robin Schot*,[1], Imme Schuringa*,[1], Álvaro Rodríguez Echarri[1], Lars Sonneveld[1], Tom Veeken[1], Yang Lu[2], Samuel D. Stranks[2], Albert Polman[1], Bruno Ehrler[1], and Saskia Fiedler[1]

[1]LMPV-Sustainable Energy Materials Department, AMOLF, Science Park 104, Amsterdam 1098 XG, The Netherlands

[2]Department of Chemical Engineering and Biotechnology, University of Cambridge, Cambridge, UK

## Abstract

Halide perovskite semiconductors are a promising material for high-efficiency solar cells. Their optical properties can vary within and between crystallographic grains. We present spatially-resolved cathodoluminescence (CL) spectroscopy at 2 keV and 5 keV on polycrystalline CsPbBr$_3$ perovskite films to study these variations at the nanoscale. The CL maps show a strongly reduced intensity near the polycrystalline grain boundaries. We perform numerical simulations of the far-field emission of the electron beam-generated optical near fields using the surface profiles from AFM as input. We find that near grain boundaries the light outcoupling is strongly reduced due to enhanced internal reflection and light trapping at the curved surfaces. Lateral variations in CL intensity inside grains are due to Fabry-Perot-like resonances in the film, with the substrate acting as a back reflector. Our results show that near-field coupling and interference effects can dominate nanoscale luminescence maps of halide perovskite films. The results are broadly relevant for the analysis of cathodoluminescence and photoluminescence of corrugated thin films.

## TOC Graphic

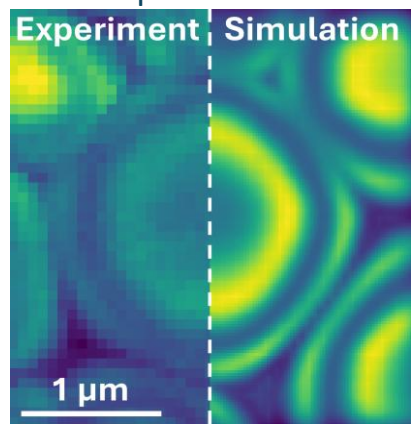

## Introduction

Halide perovskite semiconductors have emerged as a promising material for efficient solar cells with conversion efficiencies above 26%.[1] Perovskite solar cells are fabricated from polycrystalline thin films made from grains a few hundred nanometers to micrometers in size. To further increase the conversion efficiency and stability of perovskite devices, a detailed understanding of the influence of this microstructure on optoelectronic properties such as the bandgap, carrier lifetime, and recombination processes is required. In particular, measurements near grain boundaries are important, as these sites are known to act as non-radiative recombination sites, limiting optoelectronic device performance.[2,3]

Cathodoluminescence (CL) spectroscopy is a well-established technique to study the optoelectronic properties of semiconductors at high spatial resolution. In CL, a high-energy electron beam is raster-scanned over the surface, generating a cascade of inelastic scattering events inside the material. The electron beam generates a dense cloud of excited charge carriers, which then diffuse and recombine, the latter either by radiative emission (i.e., resulting in CL) or non-radiative recombination. Previous CL studies on perovskite thin films have helped identify halide segregation,[4] which results in spatial variations of the bandgap.[5] Also, variations in the CL signal have been observed between different crystal grains, indicating that the structure of the perovskite films can be laterally inhomogeneous.[3,6] Time-resolved CL has been used to probe the sub-nanosecond decay of the charge carrier population in $CsPbBr_3$ perovskite films, which was attributed to the fast diffusion of carriers away from the small electron-sample interaction volume.[7] Similarly, in other photovoltaic semiconductors, such as kesterite thin films, CL spectroscopy has proven to be a powerful technique to study inhomogeneities in materials quality.[8,9]

Spatially-resolved CL maps represent the CL intensity measured at each position of the electron beam on the sample. However, on the corrugated surfaces common in halide perovskite films, the outcoupling of CL emission can be strongly influenced by the surface. In particular, transmission through the surface interface must satisfy electromagnetic boundary conditions given by Maxwell's equations that depend on the orientation of the electromagnetic fields inside the semiconductor relative to the surface. As CL is generated in a volume smaller than the wavelength of light, the CL outcoupling cannot be described by geometric ray optics but is governed by radiation from optical near fields. Therefore, the evanescent electromagnetic near-field components must be considered in analyzing the CL maps.

In this work, we perform spatially-resolved CL spectroscopy on polycrystalline $CsPbBr_3$ perovskite films. We find that the collected CL intensity is strongly reduced for electron beam positions near grain boundaries. Comparing the CL maps with atomic force microscopy (AFM) maps on the same area and using numerical simulations of the light outcoupling, we find that reduced intensity at grain boundaries is directly related to their morphology. Reduced outcoupling of the CL emission at these points is due to internal reflection of evanescent optical near fields. We also find that measured lateral CL intensity variations across the perovskite grains are due to the interference of light reflected at the film top surface and substrate interface. These spatial variations, therefore, need to be taken into account when interpreting the spatially-resolved CL emission from perovskite films.

To study the correlation between microstructure and the CL signal, we use a thermally co-evaporated polycrystalline $CsPbBr_3$ film on silicon with a nominal thickness of 300 nm (see Methods). This composition is relatively stable under e-beam irradiation, making it ideal for isolating optoelectronic effects with minimal beam degradation.[10–12] Characteristic X-ray diffractograms (XRD) for both unannealed and annealed samples are shown in Figure S1 in the Supplementary Information (SI), indicating that CsPbBr$_3$ is in the orthorhombic phase.[13–15] We use different annealing protocols to enhance the crystallinity of the films and create films with different grain sizes.

Figure 1(a-c) shows the atomic force microscopy (AFM), secondary electron (SE), and polychromatic CL maps of the same area of the CsPbBr$_3$ perovskite film. The SE image and CL map were simultaneously acquired with an acceleration voltage of 5 kV. The AFM map and SE image reveal a structure with grain sizes ranging from ~100 nm to ~1 μm. The AFM map shows surface height variations of up to 120 nm across the surface topography. A characteristic CL spectrum of a grain interior (single pixel) is shown in Figure 1(d), revealing an emission peak around λ = 525 ± 3 nm (~2.36 eV) at the bandgap of CsPbBr$_3$.[16] The dark, optically inactive grains observed in the CL map can be attributed to perovskite in the CsPb$_2$Br$_5$ phase, as evidenced by the XRD diffraction patterns in Figure S1.[14,15,17] Otherwise, the peak CL intensity within grains is relatively consistent across the image, suggesting good homogeneity in the optical properties of the perovskite film across different grains.

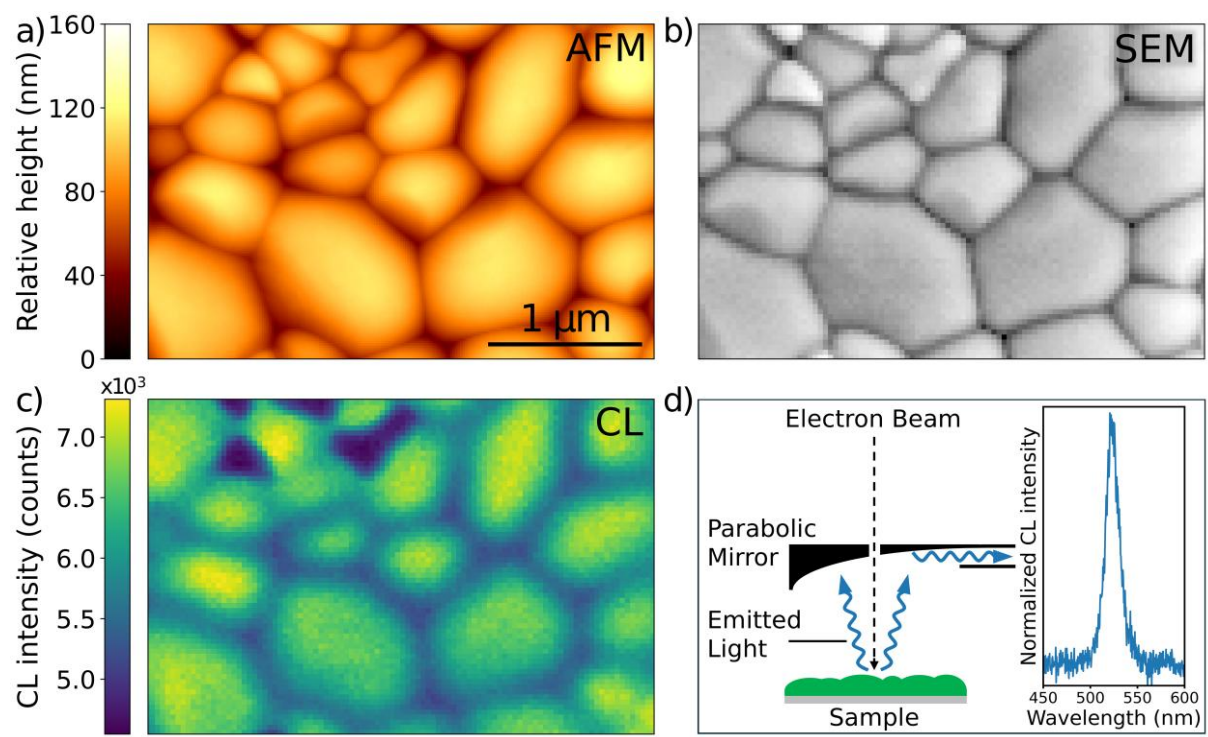

Figure 1. Surface analysis of CsPbBr$_3$ perovskite thin film. (a) AFM topography, (b) 5 keV secondary electron (SE) image, (c) 5 keV CL map (λ = 525 nm, 6 nm bandwidth), and (d) schematic of CL measurement geometry and CL spectrum taken from one of the perovskite grains. The scale bar in (c) applies to the three maps.

Notably, however, the CL map shows a reduced intensity at the grain boundaries. Often, reduced emission efficiency at grain boundaries is assigned to non-radiative recombination losses.[3] However, the strong morphological changes should affect the outcoupling of emitted light.[18] This consideration raises the question as to whether the change in intensity at the grain boundaries is due to reduced CL radiation quantum efficiency near the grain boundaries or due to an optical effect resulting from variations in outcoupling of the CL emission from the film.

To study the effect of surface topography on the outcoupling of CL emission, we analyze data for electron energies of 2 keV and 5 keV. Their corresponding electron-sample interaction volumes differ significantly, with the 5 keV interaction volume extending deeper into the perovskite compared with the 2 keV interaction volume. Figures 2(a,b) show Monte Carlo simulations of primary electron trajectories.[19] The contour lines represent the fraction of the energy density deposited in the film by the cascade of inelastic electron collisions; the yellow and purple regions indicate high-energy and low-energy secondary electrons, respectively. The 2 keV electrons deposit most of their energy in a ~30-nm-shallow region beneath the surface, while the 5 keV electrons deposit energy up to around ~150 nm. Similarly, the radial distribution of energy deposition is much wider for 5 keV than that of 2 keV irradiation. If the morphology significantly affects CL outcoupling, intensity variations should be more pronounced at 2 keV than at 5 keV, while non-radiative recombination effects remain comparable.

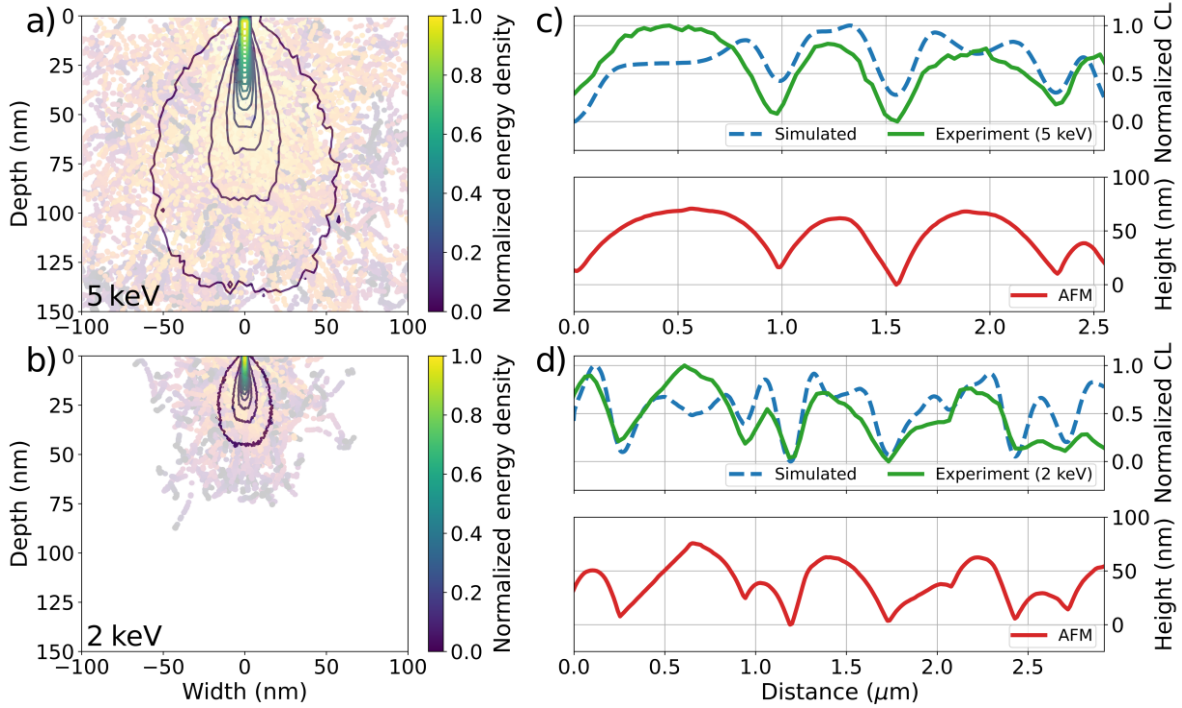

Figure 2. (a,b) Monte Carlo simulations of electron cascades in a CsPbBr$_3$ perovskite thin film at (a) 5 keV and (b) 2 keV electron energy.[19] Dots indicate inelastic collisions and yellow and purple dots indicate high- and low-energy electrons, respectively. The fraction of the energy density deposited in inelastic collisions is indicated by the contour lines. (c,d) Experimental AFM (red) and normalized CL (green) line traces ($\lambda$ = 525 nm, 6 nm bandwidth) across the perovskite film on silicon at (c) 5 keV and (d) 2 keV, shown together with simulated CL line traces (blue dashed lines). The silicon substrate was positioned 171 nm and 151 nm below the AFM minimum for the 5 keV and 2 keV case, respectively, as derived from cross-sectional SEM measurements in Figure S3 in the SI. The experimental CL line trace is plotted with an offset in the x-direction to match best with the dips in the grain boundaries in the simulated trace.

To calculate the CL emission from the corrugated surface topographies, we perform two-dimensional numerical finite-difference time-domain (FDTD) simulations of the electromagnetic near-field and far-field distribution for an array of dipole emitters placed inside the perovskite film. AFM line traces are used as an input for the simulations, and the simulated dipole distribution in the film follows the AFM profile conformally. The far-field emission for each dipole position corresponds to the collected CL intensity at that position. We calculate the incoherent sum of emission from an array of dipoles (summing contributions of x-, y-, and z-orientated dipoles) inside the interaction volume, weighted by the deposited energy density derived from the Monte Carlo simulations (for details, see Methods). We then calculate the total angle-integrated far-field radiation intensity to reflect the CL intensity that is collected over the upper hemisphere. Note that we do not consider diffusion of the charge carriers before emission in our simulations, i.e., we assume that the emission dipoles are located at the position of excitation with the electron beam.

Figure 2(c) displays the experimental AFM and the normalized CL intensity at the CsPbBr₃ bandgap (λ = 525 nm, 6 nm bandwidth) at a line trace taken across the maps in Figure 1(a,c) for 5 keV irradiation. Similar data for 2 keV are shown in Figure 2(d). The positions of these traces within the two-dimensional CL maps are shown in Figure S2 in the SI. For both line traces, the AFM profile across the film shows valleys in the surface topography at the position of the grain boundaries. Comparing the AFM topographies with the CL profiles, a clear correlation between the topographic valleys and CL intensity minima is observed.

Figures 2(c,d) also show the line profiles of the CL emission calculated using the FDTD procedure described above. The overall profile of the calculated CL intensity matches well with the experimental CL line trace. This indicates that the film morphology plays a significant role in determining the CL outcoupling efficiency as no non-radiative recombination was considered in our simulations. The simulations show that the curved surface near grain boundaries enhances internal reflection of light, resulting in increased light trapping within the perovskite film. As a result, when the electron beam is placed near a grain boundary, the outcoupled CL intensity is significantly reduced.

Figure 3(a) shows the CL map (excitation 2 keV, emission λ = 525 nm, 6 nm bandwidth) for a CsPbBr₃ film with larger grains. The grain sizes range from approximately 1 to 4 µm, which was achieved by annealing the films at 200 °C for 30 minutes (see Methods). Here, in addition to the dark bands at the grain boundaries, similar to those seen in Figure 2(c,d), we observe pronounced CL intensity modulations within individual grains. These intragrain patterns show concentric rings with alternating emission intensity for nearly all perovskite grains. Figure 3(b) presents a line trace from the experimental CL map overlayed with a simulation using the AFM profile as input. Just like before, the low CL signal near the grain boundaries is well represented by the simulations, reflecting the reduced outcoupling of CL emission. At the same time, both the experiment and the simulations show a ring of reduced emission near the middle of the grain, with a slight enhancement at its center. This intragrain variation is particularly clear for the large grain represented by the AFM scan between 3-4 µm in Figure 3(b,c).

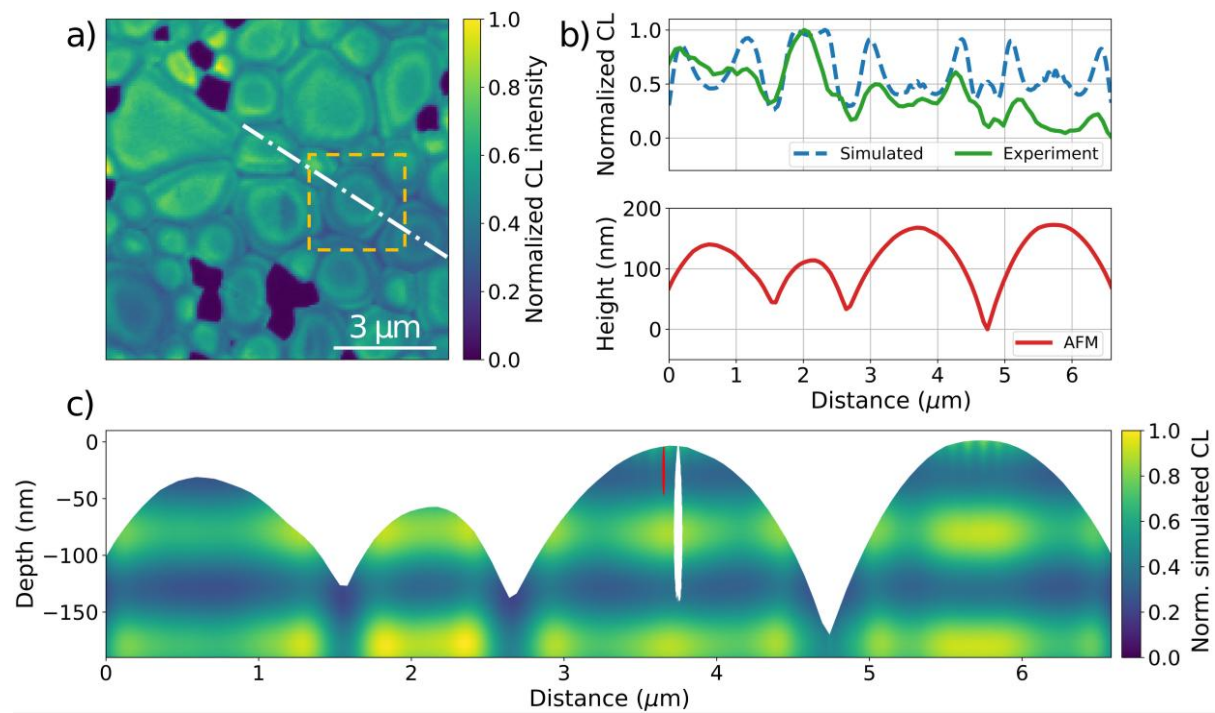

Figure 3. (a) Experimental CL map (2 keV, λ = 525 nm, 6 nm bandwidth) for a CsPbBr₃ film with grain sizes ranging from ~1 to 4 µm. The white dashed line indicates the line trace for the 2D simulations.

The dashed orange square indicates the area that has been simulated in 3D. (b) AFM (red) and experimental CL line traces (green) across the perovskite film together with the simulated CL line trace (dashed blue line). The experimental line trace is plotted with an offset in the x-direction to coincide with the simulated trace at the grain boundaries. (c) Total integrated far-field radiation versus dipole position in the perovskite film placed on a silicon substrate positioned 70 nm below the AFM minimum, based on the crosscut measurements in Figure S3. The 2 keV and 5 keV interaction volumes of Figure 2(a,b) are displayed in red and white, respectively. Note the different horizontal and vertical length scales in Figure 3(c).

To study the origin of the intragrain CL modulation in more detail, Figure 3(c) presents a crosscut through perovskite grains, using the surface profile from the AFM line trace of Figure 3(b). The color scale indicates the simulated angle-integrated far-field radiation intensity for each dipole position within the perovskite film. The simulations show a clear dependence of the outcoupled CL on the dipole position within the grain. First, the dark regions corresponding to the dark CL bands at the grain boundaries are clearly seen again. Furthermore, outcoupling variations are observed depending on the dipole position inside the grains, both horizontally and vertically. We ascribe the vertical variations to constructive and destructive interference of dipole emission reflecting off the surface and the silicon substrate below the perovskite film. This reasoning agrees well with the observed fringe period of roughly 100 nm, which is consistent with the thin-film interference period for 525 nm light in a medium with a refractive index of the CsPbBr$_3$ perovskite (n=2.67).[20]

Figure 3(c) also shows the dimensions of the electron cascades for 2 keV and 5 keV electrons. The 2 keV excitation volume extends ~30 nm into the film (Figure 2b) and scanning the electron beam across the sample surface thus results in the detection of CL from near-surface regions with intensities that reflect the constructive or destructive interference in the film. This surface-sensitive detection gives rise to the observation of dark and bright rings in the CL map of Figure 3(a). For the 5 keV beam, which penetrates deeper into the film (exceeding 100 nm in depth), the collected CL emission is an average over multiple interference minima and maxima, causing the ring structure to be less pronounced. This phenomenon agrees with the 5 keV CL map in Figure S4 in the SI, in which the rings are less pronounced compared with those in the 2 keV CL map in Figure 3(a).

To further demonstrate that the observed rings originate from interference, we semi-analytically calculate the total far-field radiation for a two-dimensional array of dipoles embedded in a planar thin film of perovskite between silicon and air (Figure S8 in the SI). The calculated far-field intensity (Figure S9(a) in the SI) is weighted with the laterally-averaged excitation density in the Monte Carlo cascade profiles for 2 and 5 keV electrons (Figure S9(b)), from which we derive the CL emission intensity as a function of dipole depth (Figure S9(c)). We then calculate the weighted far-field CL intensity integrated over depth for different perovskite thicknesses for 2 keV and 5 keV electrons (Figure S10). This clearly shows the variation of the outcoupled CL intensity with film thickness, confirming the interference model. The lower visibility of the interference rings for 5 keV irradiation that is observed in experiment (Figure S4 compared to Figure 3(a)) is also confirmed by the calculation in Figure S10.

So far, we have presented analysis for arrays of dipole emitters in the x-z plane, which intersects the line traces in the AFM and CL maps. Next, we study the outcoupling effects in a geometry that extends into the third dimension inside the film. Figure 4(a) displays the AFM topography (top layer) for the grain within the orange dashed square in Figure 3(c). We simulate the far-field CL intensity for a sheet of dipoles which conformally follows the surface topography at a fixed depth. Data are shown for depths of 10, 20, 30, 40, and 50 nm. At each depth the concentric rings in the simulated CL emission intensities are clearly visible. Their intensity maxima shift inward as the dipole layer is positioned deeper beneath the surface, as expected for the interference model given the convexly curved surface of the grain (section S7 in the SI). The simulated CL map for a dipole depth of 30 nm is overlaid on the experimental CL map for 2 keV irradiation in Figure 4(b). It matches very well with the experimental data, in agreement with the fact that 2 keV electrons deposit most of their energy around that depth. This is confirmed by calculations in Figure S11 of the SI, that show the depth of maximum CL emission for 2 keV electrons is 20-30 nm, depending on the perovskite film thicknesses.

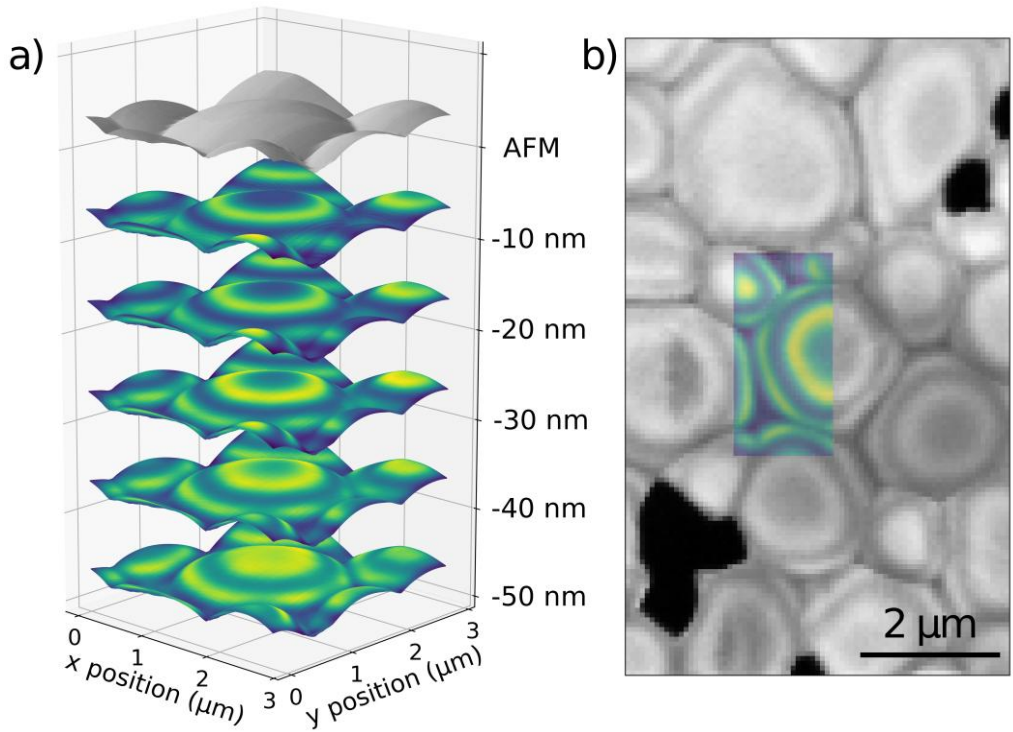

Figure 4. (a) Top: AFM map of a large perovskite grain. Colored patterns: simulated integrated far-field CL intensity for an array of dipoles in a sheet conformally following the surface topography positioned at depths of 10, 20, 30, 40, and 50 nm. (b) Experimental CL map (2 keV, λ = 525 nm, bandwidth 6 nm, grey scale) for a CsPbBr$_3$ film with grain sizes ranging from ~1 to 4 μm, with the center grain shown in the AFM map in (a). Overlayed colored region: simulated far-field CL intensity for a sheet of dipoles at 30 nm depth as shown in (a).

Overall, our analyses show that the trends measured in the CL maps are mostly determined by optical outcoupling variations and interference effects. In the analyses we have assumed unity internal quantum efficiency (IQE = 1), meaning that all energy used in the excitation of dipoles is released in the form of radiation. In practice, the IQE may be well below unity, in which case the emission intensity collected from dipoles at a certain position is modulated by the Purcell factor, which is proportional to the local density of optical states (LDOS) (section S5 in the SI). Indeed, due to the strong index contrast between the perovskite and vacuum and the presence of a high-index substrate, the Purcell factor is modulated across the film.[21–24]

We have simulated the Purcell factor as a function of position in the film, alongside the dipole simulations shown above (Figures S5(b,c,d) and S6(b,c,d) in the SI). We then multiplied the outcoupling maps shown in Figure 2(c,d) with this Purcell factor distribution to represent the limit of IQE << 1 (Figures S5(a) and S6(a) in the SI). We find only a minor effect on the outcoupling line traces, indicating that our analysis is valid for the full range 0 < IQE < 1.

To evaluate the IQE, we have performed time-resolved CL measurements on the perovskite films and found an instrument-limited CL decay time less than 80 ps at the bandgap emission wavelength λ =

525 nm (Figure S7(a) in the SI). Time-resolved photoluminescence (PL) measurements on the same sample area on which CL maps were taken showed a PL lifetime in the nanosecond range (Figure S7(b,c)). This difference indicates that the CL measurements are taken in a regime dominated by nonradiative recombination, either due to Auger recombination caused by the high carrier injection rate in the electron cascade, or due to rapid carrier diffusion away from the small electron–sample interaction volume.[7] This implies that PL maps taken on the same areas as studied above may show (in addition to strong outcoupling effects near the grain boundaries) effects of non-radiative decay at grain boundaries that are obscured in the CL measurements due to the high bulk decay rate under electron irradiation. The low IQE in the present study therefore enables us to independently determine the optical outcoupling effects.

Finally, we address discrepancies observed between the CL measurements and the simulations, as well as the impact of certain assumptions made in the simulations. First, we note that the simulated CL traces show a stronger modulation across the surface of a grain than the measurements. This discrepancy could arise from the fact that the simulations use a two-dimensional geometry whereas in reality the three-dimensional surface topography should be taken into account. Small differences in perovskite film thickness between simulation and experiment could also influence the simulation outcome. Moreover, the Monte Carlo simulations do not capture all details of the electron-matter energy exchange[25] and carrier diffusion inside and away from the cascade may increase the effective volume from which the CL emission is generated, reducing the CL intensity modulation across the film.

Furthermore, the simulations in Figures 2-4 are carried out using a real valued refractive index, neglecting optical absorption in the perovskite film. The optical absorption length in the perovskite file is $L_\alpha = 267$ nm at the studied wavelength $\lambda = 525$ nm. Therefore, the effect of absorption on the direct outcoupling of the CL emission will be minimal for the shallow interaction depth at 2 keV (< 50 nm) (Figure S9(c) in the SI). For the 5 keV measurements, the interaction volume extends deeper, and therefore part of the CL emission that is generated deeper in the film will be absorbed, lowering the outcoupled intensity for dipoles generated at larger depth. Absorption in the perovskite does affect the visibility of the interference fringes due to the Fabry-Perot resonance across the film, as illustrated in Figure S10. Finally, we note that in the limit of high non-radiative recombination, the effect of photon recycling can be neglected. We further note that the observation of these interference effects in the 2 keV CL maps indicates that long-range carrier diffusion within the perovskite grains does not dominate the CL signal.

In conclusion, we find that intensity variations in cathodoluminescence (CL) maps of polycrystalline CsPbBr$_3$ films can be predominantly assigned to changes in outcoupling and interference effects arising from the surface morphology. This finding is in stark contrast to the conventional assumption that variations in local emission intensity stem from changes in the non-radiative recombination and therefore emission efficiency. Our analysis includes the 2D and 3D excitation distribution in the electron cascade, the effect of surface geometry on outcoupling of the electron-generated electromagnetic near fields, and the local density of optical states. More broadly, the strong near-field coupling and interference effects observed here are relevant for the analysis of CL and micro-photoluminescence of corrugated films across a wide range of material systems.

# Experimental Section

*Perovskite film synthesis*

The Si substrates were loaded to a PEROevap (CreaPhys) chamber inside a N2-filled glovebox for perovskite evaporation. Two sources, CsBr and PbBr$_2$, were both deposited at 0.25 Å/s to reach a Cs:Pb=1:1 stoichiometric ratio, which was achieved by independently controlling its own heating unit and quartz crystal microbalance rate sensor. The distance between sources to substrate holder is around 35 cm. The deposition was done at $2 \times 10^{-6}$ mbar and room temperature for the substrate. After deposition, the samples were annealed on a hotplate for 30 minutes at 200 °C in air for the films with the smaller grains (0.1-1 μm), and the samples were annealed for 5 min at 300 °C in air for the large grain films (1-4 μm).

*CL and PL measurements*

The CL measurements were performed in a FEI Quanta650 SEM (Thermo Fisher) equipped with a Delmic Sparc CL system consisting of a parabolic mirror to collect the emitted light from the sample within an angular range of 1.42 πsr, and a spectrometer with a Andor Newton CCD camera for spectral analysis. For the small-grain samples, an electron beam current of 40 pA was applied, whereas a current of 85 pA was used for the larger grains. The acceleration voltage was varied between 2 and 5 kV to enable different electron-sample interaction volumes. CL maps were recorded with a pixel size of 30 nm at an exposure time of 1 ms for each pixel for the small-grain samples whereas a pixel size of 70 nm and an exposure time of 10 ms were used for the films with larger grains. For CL lifetime measurements the electron beam was modulated to create 30 ps electron pulses using an electrostatic beam blanker at a repetition rate of 10 MHz. For PL decay time measurements were performed using a λ = 485 nm pulsed laser (3 MHz, 0.41 μW) focused with a spot size that is close to the diffraction limit. Both CL and PL decay traces were recorded using time-correlated single-photon counting systems.

*AFM measurements*

For the small-grain films, topographical maps were measured on a Bruker Dimension Icon AFM with ScanAsyst-Air probe with a nominal tip radius of 2 nm (Bruker). For the large-grain films, topographical maps were obtained with a Bruker Dimension FastScan AFM with ScanAsyst-Air probe with a nominal tip radius of 2 nm (Bruker).

*Numerical optical simulations*

Finite-Different Time-Domain (FDTD) simulations were conducted using Lumerical FDTD (Ansys), version 2024 R2.3, using surface topographies from AFM as input. Optical constants for CsPbBr$_3$ (n = 2.67) and Si (n = 4.18) were obtained from Ermoloav et al. [26] and Palik, [27] respectively. All materials in the simulations are assumed to be lossless, so the imaginary part of the refractive index was set to zero. For 2D simulations, dipole sources emitting at λ = 525 nm were positioned within the perovskite on a square mesh, with 2 nm spacing for the 2 keV traces and 5 nm spacing for the 5 keV trace. To prevent numerical, mesh-related artifacts at the surface, the first dipole line in the 5 keV simulations and the first three dipole lines in the 2 keV simulations were excluded in the analysis. The simulations for 5 keV and the 2 keV large-grain simulations extended to 200 nm beneath the surface, while the 2 keV small-grain simulations extended to 60 nm only. Three polarization directions (x, y, and z) were simulated separately, and the electric and magnetic field vectors were calculated by monitors surrounding the dipole. The simulation time was 1000 fs, and eight perfectly matched boundary layers were used to limit the simulation volume. The near-field distribution was transformed to far-field radiation fields using the open-source software package RETOP. [28] The far fields for each polarization were then incoherently summed to yield the far-field radiation intensity as a function of angle. Integrating over the full upper hemisphere resulted in the total integrated far-field radiation. Once the far-field light intensity distribution was computed for each dipole position, it was convoluted with the 2D energy deposition density of inelastic scattering for 2 keV and 5 keV electrons, using electron cascade calculations using Monte Carlo simulations using CASINO. [19] The 3D-simulations were performed using the Snellius supercomputer, hosted by SURF (the collaborative ICT organization for Dutch education and research). In all simulations, dipoles were spaced by 20 nm laterally and each plane was separated 10 nm in height. Here, because of computational limitations, the transformation to the far field was done directly in the Lumerical FDTD simulation by placing a field monitor above the surface at a fixed height. Again, the fields corresponding to each polarization were incoherently summed and integrated over the full upper hemisphere, yielding the total integrated far-field radiation.

## Author contributions

S.F., A.P., and B.E. conceived the project. Y.L. and S.D.S. fabricated the perovskite films. S.F and I.S. carried out and analyzed the CL measurements, R.S performed numerical simulations with advice from T.V.. S.F. and R.S. analyzed the simulation data. L.S. conducted the AFM measurements. A.R.E. performed analytical calculations. S.F conducted the time-resolved CL measurements. S.F. and R.S. drafted the original manuscript. A.P and B.E. supervised the project, and all authors reviewed and edited the manuscript.

## Conflicts of interest

The authors declare the following competing financial interest: Albert Polman is cofounder and co-owner of Delmic BV, a company that produces commercial cathodoluminescence systems like the one that was used in this work.

## Acknowledgments

The authors acknowledge Daphne Dekker for conducting the time-resolved PL measurements. This work is part of the research program of the Dutch Research Counsel NWO and SolarLab within SolarNL, a national research, innovation and industrial development program funded by the Netherlands National Growth Fund. It is financed by the European Research Council (ERC) under Grant Agreement No. 101019932 (QEWS), the European Innovation Counsel (EIC) under Grant Agreements No. 101017720 (EBEAM), No. 101151994 (EXPLEIN), and No. 947221 (SHAPE) under the European Union's Horizon 2020 Research and Innovation Program, and the Engineering and Physical Sciences Research Council (EPSRC) for funding (EP/S030638/1, EP/V06164X/1). S.D.S. acknowledges the Royal Society and Tata Group (UF150033, URF\R\221026). This work used the Dutch national e-infrastructure with the support of the SURF Cooperative using grant no. EINF-11575.

# Near-field effects on cathodoluminescence outcoupling in perovskite thin films
## – SUPPLEMENTARY INFORMATION –

R. Schot*,[1] I. Schuringa*,[1] A. Rodríguez Echarri,[1] L. Sonneveld,[1]
T. Veeken,[1] Y. Lu,[2] S. Stranks,[2] A. Polman,[1] B. Ehrler,[1] and S. Fiedler[1]

[1]*NWO-Institute AMOLF, Science Park 104, 1098 XG Amsterdam, The Netherlands*
[2]*Department of Chemical Engineering and Biotechnology, University of Cambridge, Cambridge, UK*

The supplementary information contains further figures complementing the discussion in the main text. Additional experimental data is shown, such as X-Ray diffractograms, line traces, a 5 keV map of the large grain sample, SEM crosscuts, and luminescence lifetime measurements. We also devoted a chapter to the effect of the non-radiative decay rate on the simulated line traces. Moreover, we present further derivations of the calculations and analytics used in a simplified one-dimensional problem. In particular, we consider that a set of dipoles is placed within an infinite parallel plate radiating to the far field. Additionally, their emission is weighted by the excitation profile provided by the incident electrons characterized by the energy deposited while penetrating the perovskite via Monte Carlo simulations.

## CONTENTS

## S1. X-RAY DIFFRACTOGRAM

Figure S1 shows the X-ray diffractogram for the perovskite films on silicon, before annealing (blue) and after annealing for 30 minutes at 200 °C (orange). The orthorhombic $CsPbBr_3$ phase is indicated with green squares, and the $CsPb_2Br_5$ phase with red circles. The appearance of the $CsPb_2Br_5$ could explain the optically inactive grains, which are visible in Figure 1 in the main text [1], [2], [3].

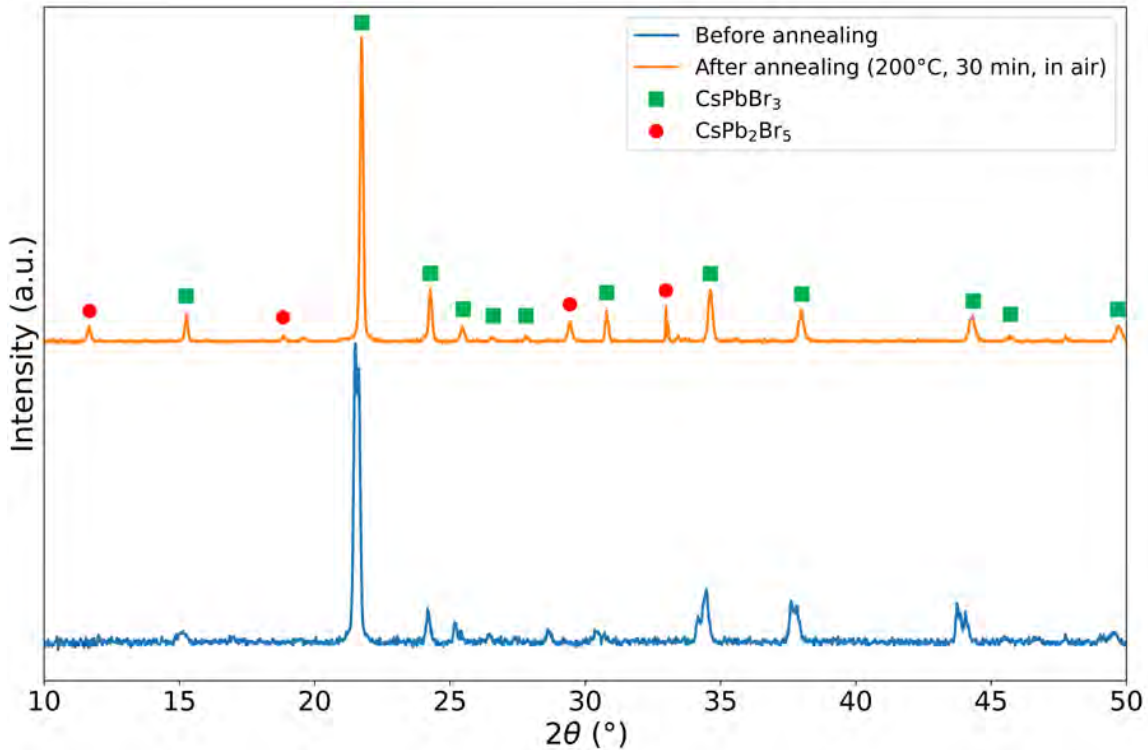

FIG. S1. X-ray diffractogram (XRD) for the perovskite films on silicon, before (blue) and after annealing (orange). The orthorhombic $CsPbBr_3$ phase is indicated with green squares, and the $CsPb_2Br_5$ phase with red circles.

## S2.   2 keV AND 5 keV AFM/CL LINE TRACES FOR THE SMALL-GRAIN SAMPLES

The atomic force microscopy (AFM) and cathodoluminescence (CL) line traces for both 5 keV and 2 keV incident electrons are shown in Figure S2. The simulations belonging to the traces are shown in Figure 2 in the main text.

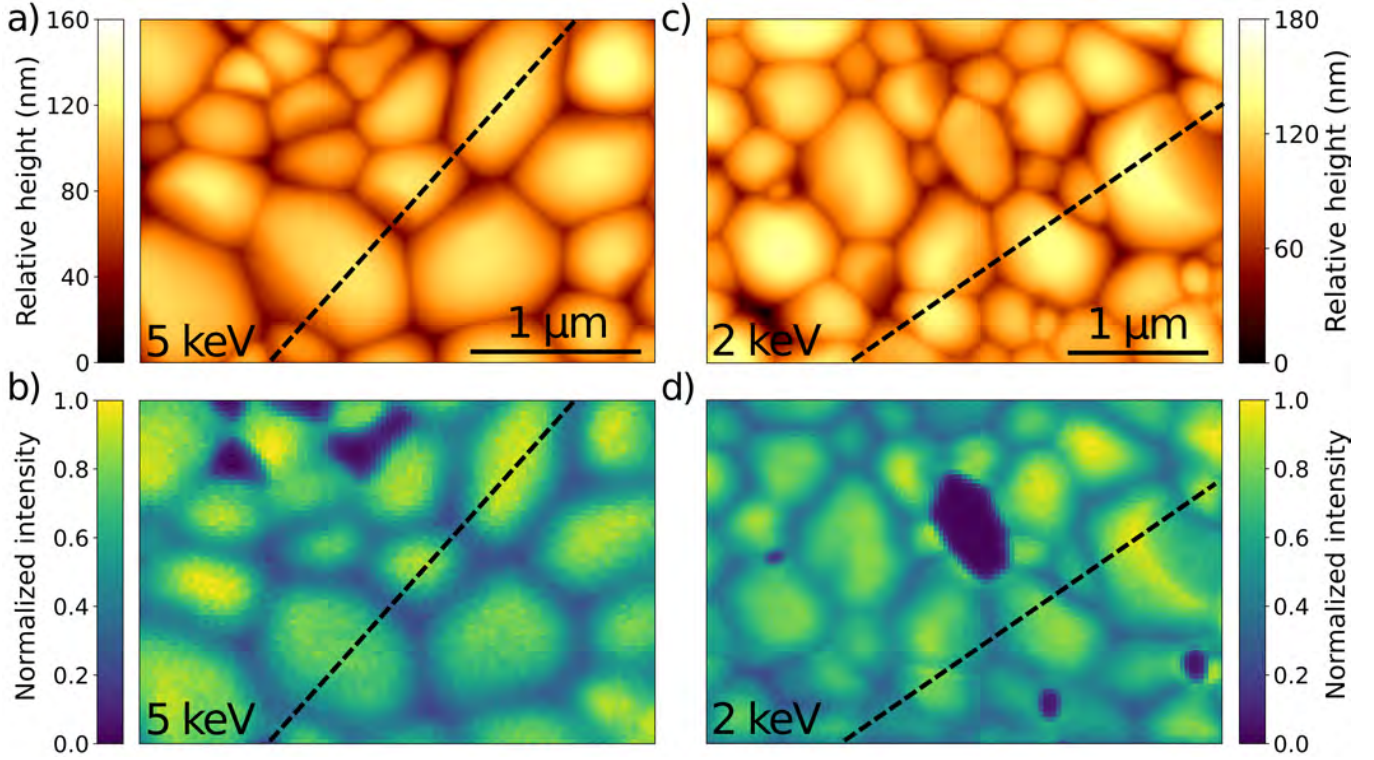

FIG. S2. AFM and CL line traces for (a,b) 5 keV and (c,d) 2 keV incident electrons, indicated with a dashed black line. The scalebar in the AFM maps applies to both the AFM and CL map.

## S3. PEROVSKITE THICKNESS MEASUREMENTS

The silicon substrate depth was calculated by subtracting the average perovskite grain thickness from the average relative grain height measured with AFM. The grain thickness was obtained from cross-sectional SEM measurements on cleaved samples, shown in Figure S3(d,e). The AFM average peak height was determined numerically by locating points where the derivative of the surface profile equaled zero. To exclude grains growing on top of other grains, a maximum height threshold of 130 nm (for the 5 keV map) and 160 nm (for the 2 keV map) was applied. These substantially higher grains are particularly evident in Figure S3(b). For the 2 and 5 keV small-grain simulations, the two AFM maps were analyzed separately, while the same set of crosscut measurements were used. Figures S3(a,b) show AFM maps with the identified peaks, along with cross sections for the small- and large-grain simulations in (d,e).

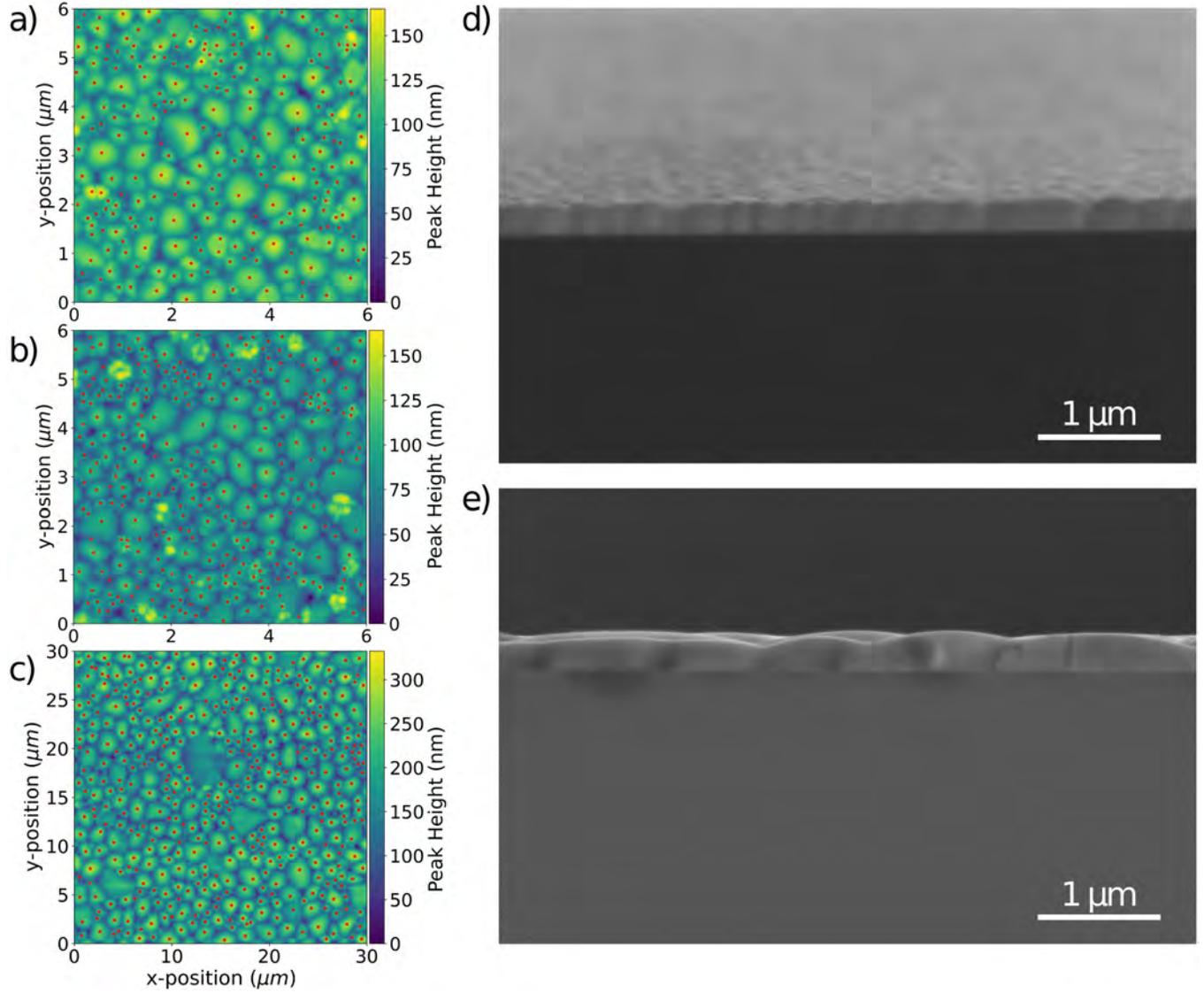

FIG. S3. AFM maps for (a) 2 keV small-grain, (b) 5 keV small-grain, and (c) 2 keV large-grain samples. Red dots indicate numerically calculated grain peak positions, where the derivative of the surface profile equaled zero. SEM crosscuts for (d) small-grain and (e) large-grain samples.

## S4. 5 keV MEASUREMENT ON LARGE-GRAIN FILM

A 5 keV CL map of the same film as the 2 keV map in Figure 3(a) is shown in Figure S4. Compared with the 2 keV map, the rings are less clearly visible, which is due to the larger excitation volume of the 5 keV electrons, as described in the main text.

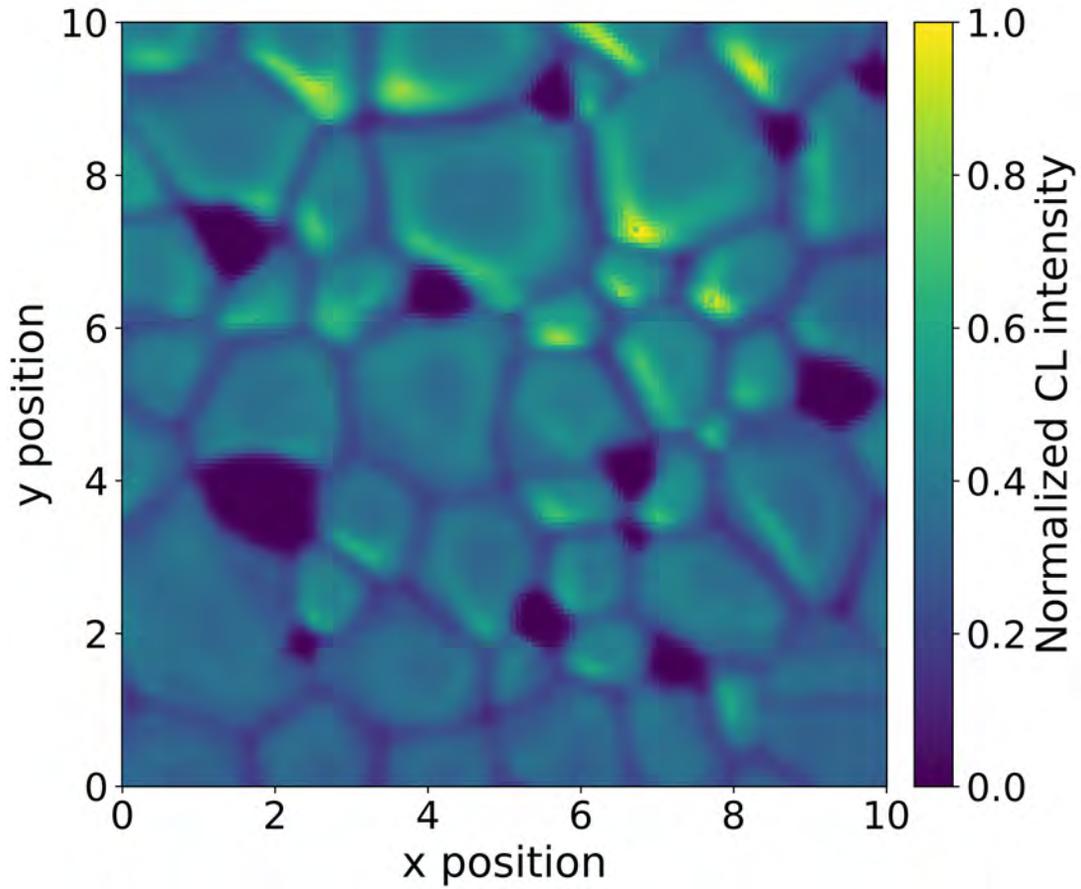

FIG. S4. Experimental CL map (5 keV, λ=525 nm, 6 nm bandwidth) for a CsPbBr₃ film with grain sizes ranging from roughly 1 to 4 μm.

## S5. THE EFFECT OF THE NON-RADIATIVE DECAY RATE

The spontaneous emission rate of an optical emitter is governed by the local optical density of states (LDOS), that reflects the density of optical modes to which the emitter can couple. Following the pioneering work of Drexhage [4] several experiments have demonstrated how the variation of the LDOS near an interface [5] or across a thin film [6], [7] results in a depth-dependent radiative rate of embedded emitters. The modification of the LDOS is often expressed in terms of the Purcell factor $F_p(\mathbf{r})$, which quantifies the enhancement or suppression of the radiative decay rate relative to that in a homogeneous reference medium. In the case of a non-unity internal quantum efficiency (IQE), variations in the LDOS lead to corresponding changes in the Purcell factor, thereby modulating the luminescence intensity. Under steady-state excitation in the linear pump regime (i.e. not in saturation), which is common for CL and PL experiments, the position dependent intensity is given by

$$\mathrm{I}(\mathbf{r}) = \eta_{\mathrm{out}}(\mathbf{r}) \, \frac{\Gamma_r(\mathbf{r})}{\Gamma_r(\mathbf{r}) + \Gamma_{nr}(\mathbf{r})} = \eta_{\mathrm{out}}(\mathbf{r}) \, \frac{F_p(\mathbf{r}) \, \Gamma_{r,0}}{F_p(\mathbf{r}) \, \Gamma_{r,0} + \Gamma_{nr}(\mathbf{r})} \tag{1}$$

with $\eta_{\mathrm{out}}$ the outcoupling efficiency, $\Gamma_r(\mathbf{r})$ the radiative decay rate, $\Gamma_{nr}(\mathbf{r})$ the non-radiative decay rate, $F_p(\mathbf{r})$ the Purcell factor, and $\Gamma_{r,0}$ the radiative decay rate of the dipole in a homogeneous reference medium. In the limit of $\Gamma_r(\mathbf{r}) \gg \Gamma_{nr}(\mathbf{r})$ (IQE = 1), this equation reduces to

$$\mathrm{I}(\mathbf{r}) \approx \eta_{\mathrm{out}}(\mathbf{r}) \, \frac{\Gamma_r(\mathbf{r})}{\Gamma_r(\mathbf{r})} = \eta_{\mathrm{out}}(\mathbf{r}) \tag{2}$$

so the collected CL or PL does not depend on the Purcell factor. In the limit of $\Gamma_r(\mathbf{r}) \ll \Gamma_{nr}(\mathbf{r})$ (IQE $\ll$ 1), equation 1 can be written as

$$\mathrm{I}(\mathbf{r}) \approx \eta_{\mathrm{out}}(\mathbf{r}) \, \frac{\Gamma_r(\mathbf{r})}{\Gamma_{nr}(\mathbf{r})} = \eta_{\mathrm{out}}(\mathbf{r}) \, F_p(\mathbf{r}) \, \frac{\Gamma_{r,0}}{\Gamma_{nr}(\mathbf{r})}. \tag{3}$$

If we assume that the non-radiative decay rate is constant throughout the film, we can write

$$\mathrm{I}(\mathbf{r}) \propto \eta_{\mathrm{out}}(\mathbf{r}) \, F_p(\mathbf{r}), \tag{4}$$

which indicates that the CL or PL intensity is directly proportional to the Purcell factor in the non-radiative regime. To investigate how the simulated CL line traces would behave in this regime, we multiplied every dipole's far field intensity with it's Purcell factor, and compared the resulting line traces with the ones simulated in the purely radiative regime (IQE = 1). The obtained 5 keV and 2 keV line traces for the small grain samples, together with the Purcell factors, are shown in Figures S5 and S6 respectively. It can be seen that the effect of the Purcell factor multiplication is negligible, indicating that variations in outcoupling efficiency are the dominant contribution to the line trace.

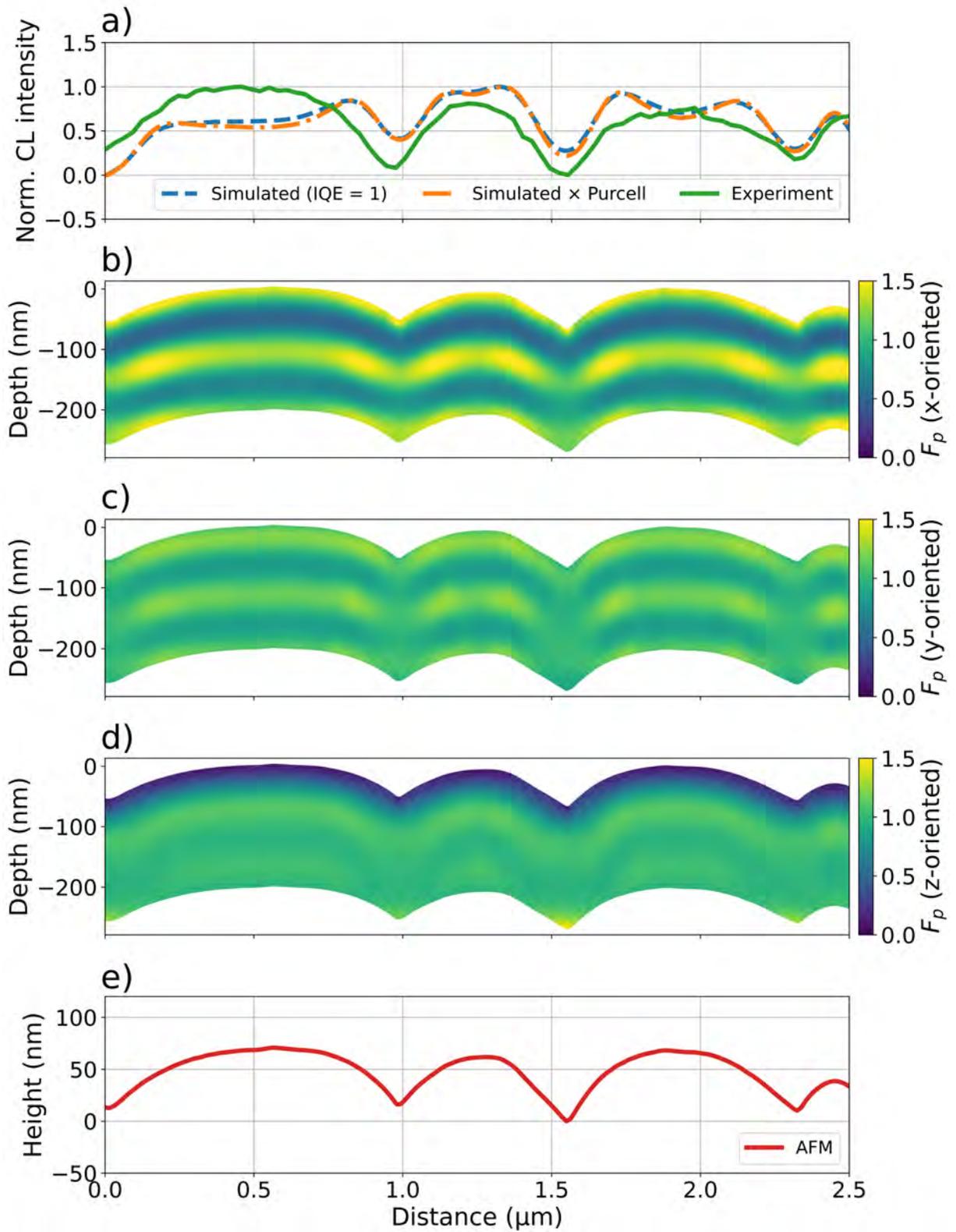

FIG. S5. (a) Experimental CL line trace (5 keV, $\lambda = 525$ nm, 6 nm bandwidth) (green) together with the simulated line traces with IQE = 1 (blue dashed line) and with IQE $\ll$ 1 (dashed orange line). (b,c,d) Simulated Purcell factors as a function of position for the same geometry as in Figure 3 (b) in the main text. (e) AFM line trace that was used as an input in the simulations.

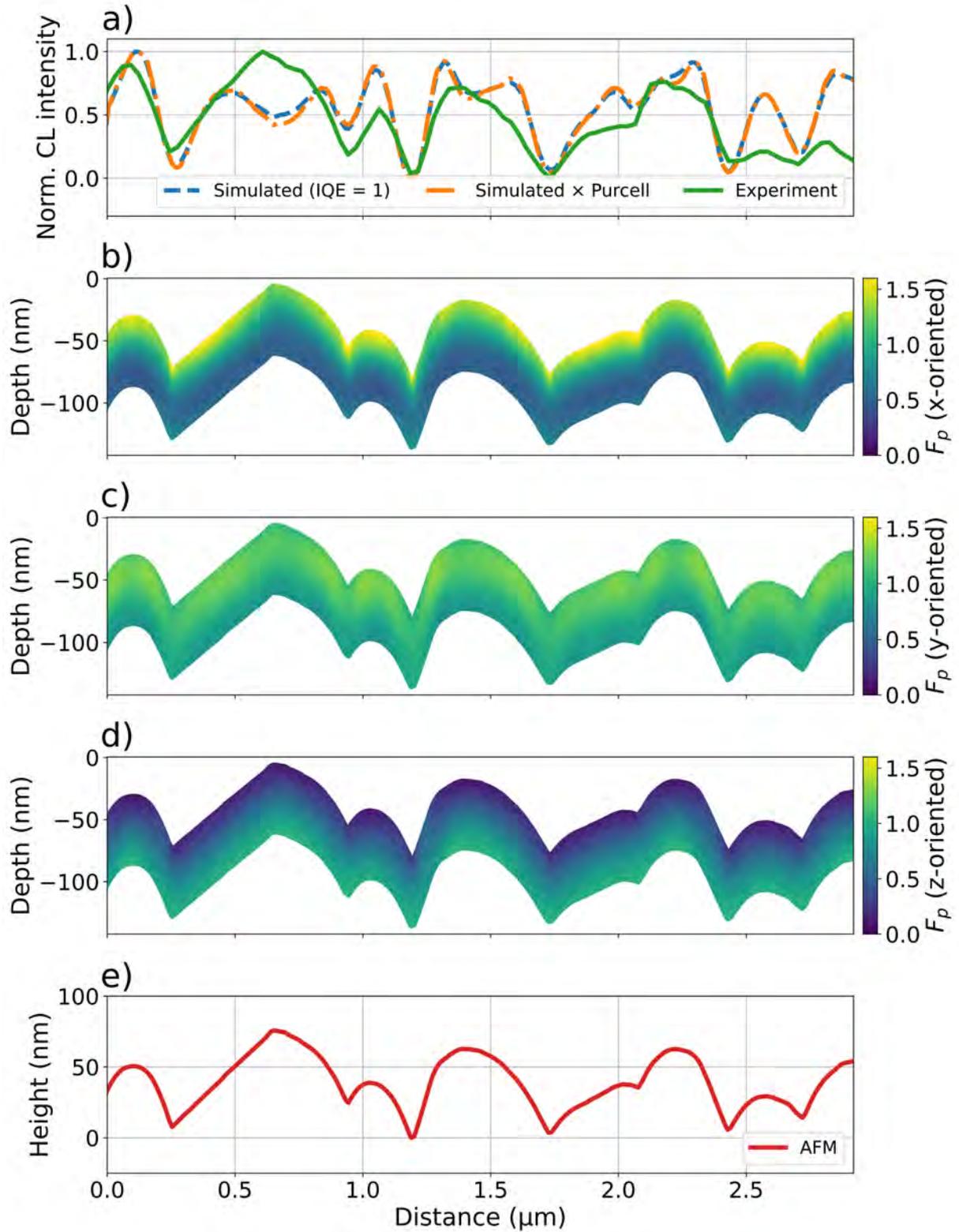

FIG. S6. (a) Experimental CL line trace (2 keV, $\lambda = 525$ nm, 6 nm bandwidth) (green) together with the simulated line traces with IQE = 1 (blue dashed line) and with IQE $\ll$ 1 (dashed orange line). (b,c,d) Simulated Purcell factors as a function of position for the same geometry as in Figure 3 (b) in the main text. (e) AFM line trace that was used as an input in the simulations.

## S6. LUMINESCENCE LIFETIMES

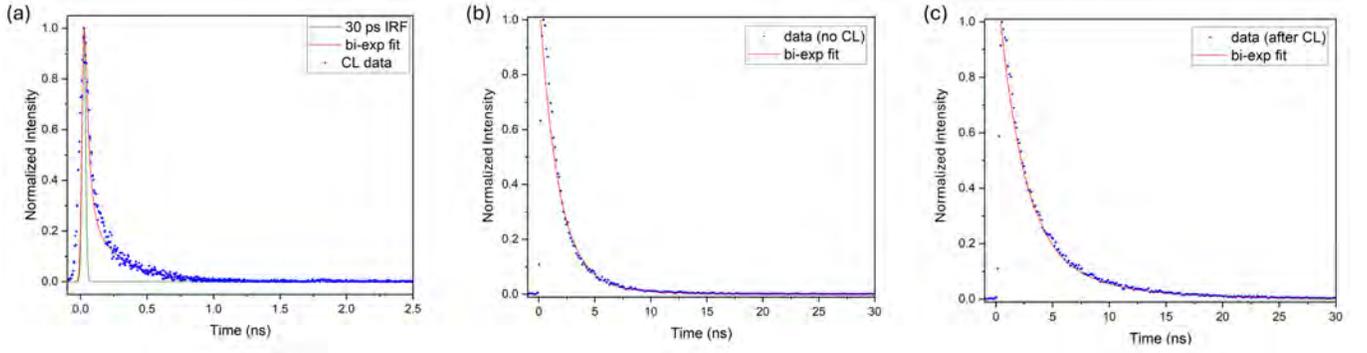

FIG. S7. (a) Time-resolved CL data (blue dots) of perovskite film with a calculated Gaussian IRF (green line) of 30 ps corresponding to the electron pulse duration of 30 ps. Bi-exponential fit (red line) with $\tau_1 = (0.03 \pm 0.01)$ ns and with $\tau_2 = (0.25 \pm 0.04)$ ns with $A_1 = (0.82 \pm 0.03)$. (b) Time-resolved PL data (blue dot) of perovskite film before electron irradiation. Bi-exponential fit (red line) with $\tau_1 = (1.6 \pm 0.2)$ ns and with $\tau_2 = (4.5 \pm 0.6)$ ns with $A_1 = (0.93 \pm 0.03)$. (c) Time-resolved PL data (blue dots) of perovskite film after electron irradiation. Bi-exponential fit (red line) with $\tau_1 = (2.2 \pm 0.1)$ ns and with $\tau_2 = (6.6 \pm 0.2)$ ns with $A_1 = (0.86 \pm 0.05)$, confirming that the electron irradiation of the sample did not result in a shortening of the dominating lifetime component ($\tau_1$).

## S7. CATHODOLUMINESCENCE OF DIPOLE EMITTERS INSIDE A THIN PLANAR FILM

To compute the energy density of radiation emitted by a set of dipoles inside a thin film, we follow a similar procedure as outlined in the supplementary material of Ref. [8]. First, we assume a dipole placed at a fixed position inside the film, accounting for the multiple internal reflections and transmission. In a second step, we derive the associated far-field and compute the fractional light emitted in the upper half-plane.

### A. Near-field of a dipole inside a slab

Let us assume a slab with permittivity $\epsilon_2$ and thickness $d$, defined from $z = 0$ to $z = d$, and placed on a substrate with permittivity $\epsilon_3$, as illustrated in Fig. S8. The space above the film is defined by $\epsilon_1$, which in general can be different from vacuum ($\epsilon \neq 1$). The electric field created by an excited dipole $\mathbf{p}$ hosted inside the thin slab of $\epsilon_2$ is given by [9, 10]

$$\mathbf{E}^{\mathrm{dip}} = \frac{1}{\epsilon_2} \left( k_2^2 + \nabla \times \nabla \right) \mathbf{p} \frac{\mathrm{e}^{\mathrm{i}k_2|\mathbf{r}-\mathbf{r}_0|}}{|\mathbf{r}-\mathbf{r}_0|}, \tag{5}$$

with $k = \omega/c$, $k_2 = k\sqrt{\epsilon_2}$, $\mathbf{r}_0 = (\mathbf{R}_0, z_0)$ is the position of the emitting dipole inside the thin slab, and $\mathbf{R}_0 = (x_0, y_0)$. Using Weyl's identity

$$\frac{\mathrm{e}^{\mathrm{i}k_2|\mathbf{r}-\mathbf{r}_0|}}{|\mathbf{r}-\mathbf{r}_0|} = \int \frac{d^2\mathbf{Q}}{(2\pi)^2} \frac{2\pi\mathrm{i}}{k_{2z}} \mathrm{e}^{\mathrm{i}\mathbf{Q}\cdot(\mathbf{R}-\mathbf{R}_0)} \mathrm{e}^{\mathrm{i}k_{2z}|z-z_0|}, \tag{6}$$

in which $\mathbf{Q} = (Q_x, Q_y)$ and $k_{2z} = \sqrt{k_2^2 - Q^2}$ (with $\mathrm{Im}\{k_{2z}\} > 0$), we then rewrite the electric field produced by a dipole inside a homogeneous medium as

$$\mathbf{E}^{\mathrm{dip}} = \frac{\mathrm{i}}{2\pi\epsilon_2} \int \frac{d^2\mathbf{Q}}{k_{2z}} \left[ k_2^2\mathbf{p} - (\mathbf{p}\cdot\mathbf{k}_2^{\pm})\mathbf{k}_2^{\pm} \right] \mathrm{e}^{\mathrm{i}\mathbf{Q}\cdot(\mathbf{R}-\mathbf{R}_0)} \mathrm{e}^{\mathrm{i}k_{2z}|z-z_0|}, \tag{7}$$

with "+" sign for $z > z_0$ and "−" when $z < z_0$, and $\mathbf{k}_2^{\pm} = \mathbf{Q} \pm k_{2z}\hat{z}$. Note that the unit vector $\hat{\mathbf{k}}_2^{\pm} = \mathbf{k}_2^{\pm}/k_2$ in combination with $\mathbf{e}_s = (-Q_y\hat{x} + Q_x\hat{y})/Q$ and $\mathbf{e}_{p,2}^{\pm} = (\pm k_{2z}\mathbf{Q} - Q^2\hat{z})/(Qk_2)$ form a complete set of orthonormal vectors, and therefore, we can write

$$\mathbf{E}^{\mathrm{dip}} = \frac{\mathrm{i}k^2}{2\pi} \int \frac{d^2\mathbf{Q}}{k_{2z}} \left[ (\mathbf{e}_s\cdot\mathbf{p})\mathbf{e}_s + (\mathbf{e}_{p,2}^{\pm}\cdot\mathbf{p})\mathbf{e}_{p,2}^{\pm} \right] \mathrm{e}^{\mathrm{i}\mathbf{Q}\cdot(\mathbf{R}-\mathbf{R}_0)+\mathrm{i}k_{2z}|z-z_0|}. \tag{8}$$

The field above the thin film ($z > d$) is the result of the electric field produced by the dipole at the position $z = z_0$ that undergoes multiple reflections inside the thin slab (i.e., like Fabry-Pérot). Thus, the resulting electric field yields

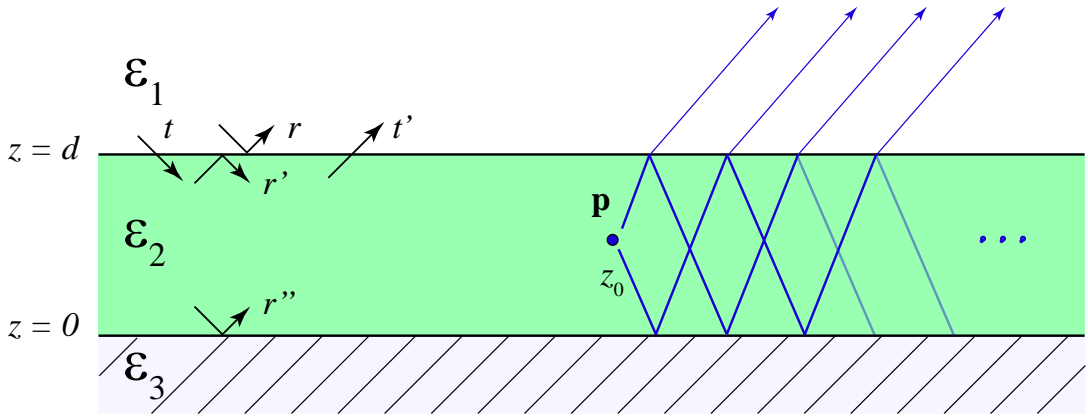

FIG. S8. **Illustration of a dipole p emitting inside a film**. A dipole radiating to the far field is placed at $z = z_0$ inside a film of thickness $d$ and permittivity $\epsilon_2$ supported by a substrate of permittivity $\epsilon_3$ and covered by cladding at the upper space defined by $\epsilon_1$. The emitted waves up- and downwards originated by the dipole suffer multiple reflections before they escape to the far field, which is described by the corresponding reflection and transmission coefficients.

$$
\mathbf{E}^{\mathrm{dip}} = \frac{\mathrm{i}k^2}{2\pi} \int \frac{d^2\mathbf{Q}}{k_{2z}} \mathrm{e}^{\mathrm{i}\mathbf{Q}\cdot(\mathbf{R}-\mathbf{R}_0)+\mathrm{i}k_z(z-d)} \left\{
\begin{array}{l}
\mathbf{e}_{p,1}^+ \left[ (\mathbf{e}_{p,2}^+\cdot\mathbf{p}) \dfrac{t_{\mathrm{p}}'\mathrm{e}^{\mathrm{i}k_{2z}(d-z_0)}}{1-r_{\mathrm{p}}'r_{\mathrm{p}}''\mathrm{e}^{2\mathrm{i}k_{2z}d}} + (\mathbf{e}_{p,2}^-\cdot\mathbf{p}) \dfrac{r_{\mathrm{p}}''t_{\mathrm{p}}'\mathrm{e}^{\mathrm{i}k_{2z}(d+z_0)}}{1-r_{\mathrm{p}}'r_{\mathrm{p}}''\mathrm{e}^{2\mathrm{i}k_{2z}d}} \right] \\[4mm]
+ \mathbf{e}_s(\mathbf{e}_s\cdot\mathbf{p}) \dfrac{t_{\mathrm{s}}'}{1-r_{\mathrm{s}}'r_{\mathrm{s}}''\mathrm{e}^{\mathrm{i}2k_{2z}d}} \left[ \mathrm{e}^{\mathrm{i}k_{2z}(d-z_0)} + r_{\mathrm{s}}''\,\mathrm{e}^{\mathrm{i}k_{2z}(d+z_0)} \right]
\end{array}
\right\}, \tag{9}
$$

where we have that the light waves have reflected multiple times inside the thin film and have already transmitted to medium 1 (i.e., right above the film). We use the well-known Fresnel coefficients [10, 11] for planar interfaces that describe how light scatters when going from medium 1 to medium 2, which are defined as

$$
r_{\mathrm{s},12} = \frac{k_{z1}-k_{z2}}{k_{z1}+k_{z2}}, \qquad\qquad t_{\mathrm{s},12} = \frac{2k_{z1}}{k_{z1}+k_{z2}}, \tag{10}
$$

$$
r_{\mathrm{p},12} = \frac{\epsilon_2 k_{z1}-\epsilon_1 k_{z2}}{\epsilon_2 k_{z1}+\epsilon_1 k_{z2}}, \qquad\qquad t_{\mathrm{p},12} = \frac{2\sqrt{\epsilon_1\epsilon_2}\,k_{z1}}{\epsilon_2 k_{z1}+\epsilon_1 k_{z2}}, \tag{11}
$$

and in particular, we used

$$
r_{\mathrm{s}}' = \frac{k_{2z}-k_{1z}}{k_{2z}+k_{1z}}, \qquad\qquad r_{\mathrm{s}}'' = \frac{k_{2z}-k_{3z}}{k_{2z}+k_{3z}}, \qquad\qquad t_{\mathrm{s}}' = \frac{2k_{2z}}{k_{2z}+k_{1z}}, \tag{12}
$$

$$
r_{\mathrm{p}}' = \frac{\epsilon_1 k_{2z}-\epsilon_2 k_{1z}}{\epsilon_1 k_{2z}+\epsilon_2 k_{1z}}, \qquad\qquad r_{\mathrm{p}}'' = \frac{\epsilon_3 k_{2z}-\epsilon_2 k_{3z}}{\epsilon_3 k_{2z}+\epsilon_2 k_{3z}}, \qquad\qquad t_{\mathrm{p}}' = \frac{2\sqrt{\epsilon_2\epsilon_1}\,k_{2z}}{\epsilon_1 k_{2z}+\epsilon_2 k_{1z}}. \tag{13}
$$

### B. Calculation of the fractional energy radiated into the far-field

Let us define as $\Gamma_{\mathrm{CL}}$ the angle-integrated Poynting vector $\mathbf{S}$ in the upper half plane (i.e., $\Omega > 0$) in the far-field per unit of frequency as

$$
\Gamma_{\mathrm{CL}}(\omega,z_0) = \frac{r^2}{\hbar\omega} \int_{\Omega>0} d\Omega \, \langle \hat{r}\cdot\mathbf{S} \rangle_{\mathrm{time}} = \frac{1}{2\pi\hbar k} \int_0^{\pi/2} \sin\theta \, d\theta \int_0^{2\pi} d\varphi \, |\mathbf{f}(\hat{r},z_0)|^2. \tag{14}
$$

In particular, the radiated energy is given by the resulting Poynting vector $\mathbf{S} = (c/4\pi)(\mathbf{E}^{\mathrm{FF}}+\mathrm{c.c.})\times(\mathbf{H}^{\mathrm{FF}}+\mathrm{c.c.})$, with the magnetic field obtained from Faraday's law $\mathbf{H}^{\mathrm{FF}} = -(\mathrm{i}/k)\nabla\times\mathbf{E}^{\mathrm{FF}}$, and the electric field defined in the far-field as $\mathbf{E}^{\mathrm{FF}} = \mathbf{f}(\hat{r},z_0)\mathrm{e}^{\mathrm{i}kr}/r$. Note that we assume a temporal dependence $\mathrm{e}^{-\mathrm{i}\omega t}$ in the magnetic and electric fields, which yields a factor of $1/2$ when performing the time average, $T^{-1}\int_0^T dt\sin^2(\omega t) = 1/2$.

To obtain the expression of $\mathbf{f}(\hat{r},z_0)$, we note that in the far-field, we take the limit of $kr\to\infty$, such that the integral in $\mathbf{Q}$ gives rise to the following type

$$
\lim_{kr\to\infty} \int \frac{d^2\mathbf{Q}}{(2\pi)^2} \mathrm{e}^{\mathrm{i}\mathbf{k}\cdot\mathbf{r}}\mathbf{g}(\mathbf{Q}) = \frac{k_z}{2\pi\mathrm{i}}\mathbf{g}(\mathbf{Q}_0)\frac{\mathrm{e}^{\mathrm{i}kr}}{r}, \tag{15}
$$

with $\mathbf{Q}_0 = (\mathbf{R}/r)k$, whose solution is found by using the stationary phase approximation (see the equation above Eq. 8.53 in Ref. [10]). Thereby, the amplitude of the electric far-field in Eq. (9) reads

$$
\mathbf{f}(\hat{r},z_0) = \mathrm{e}^{-\mathrm{i}(\mathbf{Q}_0\cdot\mathbf{R}_0+k_zd)}\frac{k^2 k_z}{k_{2z}} \left\{
\begin{array}{l}
\mathbf{e}_{p,1}^+ \left[ (\mathbf{e}_{p,2}^+\cdot\mathbf{p}) \dfrac{t_{\mathrm{p}}'\mathrm{e}^{\mathrm{i}k_{2z}(d-z_0)}}{1-r_{\mathrm{p}}'r_{\mathrm{p}}''\mathrm{e}^{2\mathrm{i}k_{2z}d}} + (\mathbf{e}_{p,2}^-\cdot\mathbf{p}) \dfrac{r_{\mathrm{p}}''t_{\mathrm{p}}'\mathrm{e}^{\mathrm{i}k_{2z}(d+z_0)}}{1-r_{\mathrm{p}}'r_{\mathrm{p}}''\mathrm{e}^{2\mathrm{i}k_{2z}d}} \right] \\[4mm]
+ \mathbf{e}_s(\mathbf{e}_s\cdot\mathbf{p}) \dfrac{t_{\mathrm{s}}'}{1-r_{\mathrm{s}}'r_{\mathrm{s}}''\mathrm{e}^{\mathrm{i}2k_{2z}d}} \left[ \mathrm{e}^{\mathrm{i}k_{2z}(d-z_0)} + r_{\mathrm{s}}''\,\mathrm{e}^{\mathrm{i}k_{2z}(d+z_0)} \right]
\end{array}
\right\}, \tag{16}
$$

where, we note that $\mathbf{k} = (\mathbf{Q}_0,k_z) = k\hat{r}$, $\mathbf{Q}_0 = k(\mathbf{R}/r) = k(\hat{x}\cos\varphi + \hat{y}\sin\varphi)\sin\theta$, and $k_z = k(z/r) = k\cos\theta$. Moreover, in the calculation of $|\mathbf{f}(\hat{r},z_0)|^2$, we obtain

$$
|\mathbf{f}(\hat{r},z_0)|^2 = \left| \frac{k^2 k_z}{k_{2z}} \right|^2 \left\{
\begin{array}{l}
\left| \dfrac{t_{\mathrm{p}}'}{1-r_{\mathrm{p}}'r_{\mathrm{p}}''\mathrm{e}^{2\mathrm{i}k_{2z}d}} \left[ (\mathbf{e}_{p,2}^+\cdot\mathbf{p})\mathrm{e}^{\mathrm{i}k_{2z}(d-z_0)} + (\mathbf{e}_{p,2}^-\cdot\mathbf{p})r_{\mathrm{p}}''\mathrm{e}^{\mathrm{i}k_{2z}(d+z_0)} \right] \right|^2 \\[4mm]
+ \left| (\mathbf{e}_s\cdot\mathbf{p}) \dfrac{t_{\mathrm{s}}'}{1-r_{\mathrm{s}}'r_{\mathrm{s}}''\mathrm{e}^{\mathrm{i}2k_2 d}} \left[ \mathrm{e}^{\mathrm{i}k_2(d-z_0)} + r_{\mathrm{s}}''\,\mathrm{e}^{\mathrm{i}k_2(d+z_0)} \right] \right|^2
\end{array}
\right\}. \tag{17}
$$

To compute the angular integrals in Eq. (14), we project the emission direction of the dipole so that

$$\mathbf{e}_s \cdot \hat{z} p_z = 0, \qquad\qquad \mathbf{e}_{p,2}^{\pm} \cdot \hat{z} p_z = \frac{-Q}{k_2} p_z, \tag{18}$$

$$\mathbf{e}_s \cdot \hat{x} p_x = \frac{-Q_y}{Q} p_x, \qquad\qquad \mathbf{e}_{p,2}^{\pm} \cdot \hat{x} p_x = \pm \frac{Q_x k_{2z}}{Q k_2} p_x, \tag{19}$$

$$\mathbf{e}_s \cdot \hat{y} p_y = \frac{Q_x}{Q} p_y, \qquad\qquad \mathbf{e}_{p,2}^{\pm} \cdot \hat{y} p_y = \pm \frac{Q_y k_{2z}}{Q k_2} p_y. \tag{20}$$

and solve for the polar angle $d\varphi$ integral, following the results

$$\int_0^{2\pi} d\varphi \left( \frac{Q_x}{Q} \right)^2 = \int_0^{2\pi} d\varphi \left( \frac{Q_y}{Q} \right)^2 = \pi, \tag{21}$$

$$\int_0^{2\pi} d\varphi \frac{Q_x Q_y}{Q^2} = \int_0^{2\pi} d\varphi \, Q_x = \int_0^{2\pi} d\varphi \, Q_y = 0, \tag{22}$$

$$\int_0^{2\pi} d\varphi = 2\pi. \tag{23}$$

After solving the above integrals, we note that the dipole emission direction can be treated independently after integrating over the azimuthal angle. Then, we write Eq. (14) for the rate of emitted photons as

$$\Gamma(\omega, z_0) = \int_0^{\pi/2} \sin\theta \, d\theta \left[ (|p_x|^2 + |p_y|^2) G_\parallel(\theta) + |p_z|^2 G_\perp(\theta) \right]. \tag{24}$$

We assume all dipoles to have the same strength $|p_x|^2 = |p_y|^2 = |p_z|^2 \equiv |p(z_0)|^2$, thereby,

$$\Gamma_{\mathrm{CL}}(\omega, z_0) = |p|^2 \int_0^{\pi/2} \sin\theta \, d\theta \left[ 2 G_\parallel(\theta) + G_\perp(\theta) \right], \tag{25}$$

where the integral can be limited to the numerical aperture (NA) of the objective if needed (here not used, see more information about it in Ref. [9]). We define the $G_\parallel$ and $G_\perp$ functions as

$$G_\parallel(\theta) = \frac{\pi}{2\pi\hbar k} \left| \frac{k_z k^2}{k_{2z}} \right|^2 \left\{ \left| \frac{k_{2z}}{k_2} \right|^2 \left| \frac{t_{\mathrm{p}}'}{1 - r_{\mathrm{p}}' r_{\mathrm{p}}'' \mathrm{e}^{2\mathrm{i} k_{2z} d}} \left[ \mathrm{e}^{\mathrm{i} k_{2z}(d - z_0)} - r_{\mathrm{p}}'' \mathrm{e}^{\mathrm{i} k_{2z}(d + z_0)} \right] \right|^2 \right.$$
$$\left. + \left| \frac{t_{\mathrm{s}}'}{1 - r_{\mathrm{s}}' r_{\mathrm{s}}'' \mathrm{e}^{2\mathrm{i} k_{2z} d}} \left[ \mathrm{e}^{\mathrm{i} k_{2z}(d - z_0)} + r_{\mathrm{s}}'' \mathrm{e}^{\mathrm{i} k_{2z}(d + z_0)} \right] \right|^2 \right\} \tag{26}$$

$$G_\perp(\theta) = \frac{2\pi}{2\pi\hbar k} \left| \frac{k_z k^2}{k_{2z}} \right|^2 \left\{ \left| \frac{Q_0}{k_2} \right|^2 \left| \frac{t_{\mathrm{p}}'}{1 - r_{\mathrm{p}}' r_{\mathrm{p}}'' \mathrm{e}^{2\mathrm{i} k_{2z} d}} \left[ \mathrm{e}^{\mathrm{i} k_{2z}(d - z_0)} + r_{\mathrm{p}}'' \mathrm{e}^{\mathrm{i} k_{2z}(d + z_0)} \right] \right|^2 \right\}, \tag{27}$$

which play a similar role like Green's functions. Remember that $Q_0 = k \sin\theta$ and $k_{jz}^2 = k_j^2 - Q_0^2$ with $k_j = (\omega/c)\sqrt{\epsilon_j}$.

### C. Cathodoluminescence intensity

The measured Cathodoluminescence intensity (CL) is given by the integration of $\Gamma_{\mathrm{CL}}(\omega, z_0)$ over all dipole positions. However, as discussed in the main text, we weight the dipole distribution by the excitation density calculated using Monte Carlo simulations. Thereby, we assume that $|\mathbf{E}^{\mathrm{ext}}(z_0)| \propto P(z_0)$, where $P(z_0)$ is the average probability of the energy deposited by the electron as a function of $z_0$.

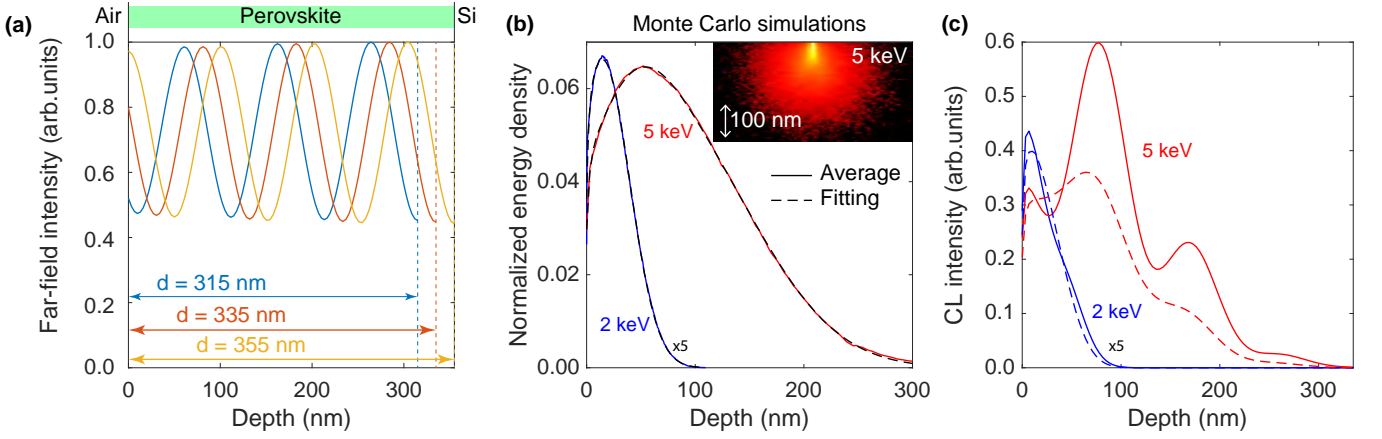

FIG. S9. **Cathodoluminescence from a thin film.** (a) Normalized far-field intensity of an excited dipole placed at $z_0$ inside a perovskite film with refractive index 2.67 at 525 nm wavelength [12]. Results are normalized to the maximum intensity for different thickness $d$ (see different colors) and computed using Eq. (24). The film is deposited on top of silicon (Si) with a refractive index of 4.18 [13], and surrounded by air (i.e., $\epsilon = 1$). (b) Averaged energy density deposited using Monte Carlo simulations in one dimension (see inset for a two-dimensional representation) by a 2-keV electron (blue solid curve) and 5-keV electron (red curve), computed using Eq. (29). Dashed lines are Gaussian fittings according to Eq. (30). (c) Resulting CL emission computed from the product of the dipole far-field emission intensity in (a) for a 335-nm thick perovskite film and the corresponding excitation profile in (b) of 2 (blue curve) and 5 keV (red) electrons. Dashed curves show the same for a lossy perovskite using $\epsilon = (2.67 + 0.1562i)^2$ [12].

In Fig. S9a, we show the calculated radiated emission intensity of a dipole into the far-field placed at position $z_0$ inside the sketched layered structure, in arbitrary units given by Eq. (24). Different colored lines indicate the emission profile for different film thicknesses (see labels). Panel Fig. S9b describes the laterally average excitation density of the electron cascade as a function of depth as described in Sec. S7 C 1 below, for 2 and 5 kV acceleration voltages (blue and red curves, respectively). Figure S9c shows the resulting CL intensity $\Gamma_{\mathrm{CL}}(\omega)$ obtained from the multiplication of the dipole emission intensity (panel Fig. S9a) with the excitation profile (panel Fig. S9b). Additionally, we compute the normalized values of $\Gamma_{\mathrm{CL}}(\omega)$ in Fig. S10a as

$$\Gamma_{\mathrm{CL}}(\omega) = \int dz_0 \, \Gamma_{\mathrm{CL}}(\omega, z_0) P(z_0), \tag{28}$$

for two different acceleration voltages and as a function of the film thickness, with damping (solid curves) and without (dashed curves). Interestingly, we observe that for the lower acceleration voltages (2 keV electrons, blue curve), the intensity varies more than at higher acceleration voltages (5 keV electrons, red curve). To quantify the contrast of the fringes, we define in Fig. S10b their visibility, respectively, showing that CL images from 2-keV electrons have a twofold resolution compared to the 5-keV ones. Note that introducing loss to the perovskite optical constant shifts the curves due to changes in the interference condition.

### 1. Fitting of energy density deposition from Monte Carlo simulations

In order to extract an effective one-dimensional energy density deposition from the three-dimensional Monte Carlo simulations $E_{\mathrm{MC}}$, we average all the probabilities through the depth of the film such that

$$P_{\mathrm{MC}}(z) = \frac{\sum_{i,j} E_{\mathrm{MC}}(x_i, y_j, z)}{\sum_{i,j} \Delta x_i \Delta y_i}, \tag{29}$$

where $\Delta x_i \Delta y_i$ is the surface element. We process the data by fitting the resulting $P_{\mathrm{MC}}(z)$ for 2 and 5 keV acceleration voltages to

$$P_{\mathrm{fitting}}(z) = \sum_{i=1}^{3} a_i \mathrm{e}^{-(z-b_i)^2/c_i^2}, \tag{30}$$

where the gaussian coefficients are in table S1.

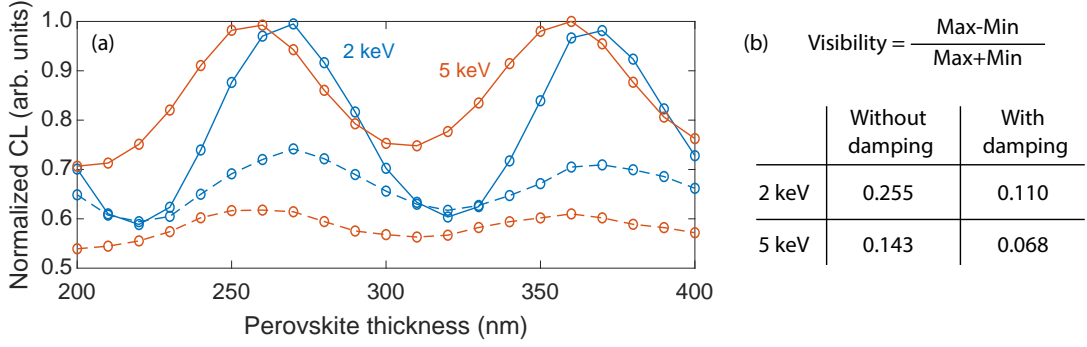

FIG. S10. **Normalized cathodoluminescence in a perovskite film.** (a) We show the normalized angle-integrated CL spectra for different perovskite film thicknesses at 2 keV (blue curve) and 5 keV (red) electrons, accounting for their electron cascade density and obtained from Eq. (28). Dashed curves simulate a lossy perovskite using $\epsilon = (2.67 + 0.1562i)^2$. (b) Calculation of the visibility using the definition given in the inset for all cases in (a).

TABLE S1. Gaussian fit coefficients for 2 kV and 5 kV acceleration voltages.

| Voltage | $a_1$ | $b_1$ | $c_1$ | $a_2$ | $b_2$ | $c_2$ | $a_3$ | $b_3$ | $c_3$ |
|---|---|---|---|---|---|---|---|---|---|
| 2 kV | $1.168 \times 10^{-4}$ | 13.02 | 1.368 | 0.01325 | 14.49 | 35.61 | $-3.514 \times 10^5$ | $-69.2$ | 16.18 |
| 5 kV | 0.01616 | 36.2 | 56.97 | 0.05166 | 74.24 | 112.7 | $-1.202 \times 10^{10}$ | $-132.8$ | 25.34 |

### D. Centroid of the dominant emission

In order to quantify at which effective depth the maximum of the emission is occurring for a given film thickness, we can compute the centroid of the emitted CL by computing

$$\text{Penetration depth centroid} = \frac{\int z_0 dz_0 \, \Gamma_{\text{CL}}(\omega, z_0) P(z_0)}{\int dz_0 \, \Gamma_{\text{CL}}(\omega, z_0) P(z_0)}, \tag{31}$$

which we show in Fig. S11 for different electron kinetic energies as a function of the perovskite thickness. We can see that the 2 keV electrons (blue curve) excite CL more effectively at a depth of 20-30 nm, depending on the thickness, while higher-energy electrons with 5 keV kinetic energy (red curve) most effectively excite CL at 90-95 nm depth. The values are calculated using 525 nm wavelength ($\sim 2.36$ eV), and while the damping does not play a major role for 2 keV electrons (dashed curves), in the case of the 5 keV electrons, the centroid drops to values around $\sim 80$ nm.

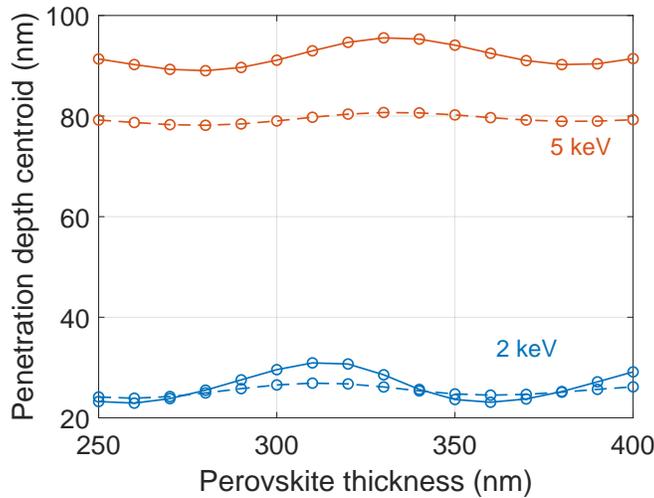

FIG. S11. **Centroid of the normalized cathodoluminescence from perovskite.** We show the centroid of the normalized integrated CL spectra for different perovskite thicknesses for 2 keV (blue curve) and 5 keV (red) electrons at 525 nm light wavelength. Solid lines do not include damping where $\epsilon = 2.67^2$, while dashed ones model the perovskite with the corresponding attenuation coefficient in which $\epsilon = (2.67 + 0.1562i)^2$.